\documentclass[aps,prl,twocolumn,showpacs]{revtex4}
\usepackage{graphicx}

\begin{document}

\title{Quantum Coherence of Image-Potential States}

\author{P. Wahl, M.A. Schneider, L. Diekh\"oner, R. Vogelgesang, and K. Kern}
\affiliation{Max-Planck-Institut f\"ur Festk\"orperforschung,
Heisenbergstr. 1, D-70569 Stuttgart, Germany}

\date{\today}

\begin{abstract}
The quantum dynamics of the two-dimensional image-potential states
in front of the Cu(100) surface is measured by scanning tunneling
microscopy (STM) and spectroscopy (STS). The dispersion relation
and the momentum resolved phase-relaxation time of the first
image-potential state are determined from the quantum interference
patterns in the local density of states (LDOS) at step edges. It
is demonstrated that the tip-induced Stark shift does not affect
the motion of the electrons parallel to the surface.
\end{abstract}


\maketitle

The image-potential states are model states for the study of
electronic interactions of electrons at surfaces, a topic that has
far reaching consequences for many surface processes. The self
interaction of electrons near metallic surfaces gives rise to
eigenstates which are confined along the surface normal by the
classical image potential on the vacuum side of the crystal
surface and by the band structure of the crystal on the other side
\cite{Echenique78}. The lifetime of electrons injected or excited
into an image-potential state is limited mainly by their
interaction with bulk electrons. This has been studied in great
detail in recent years by energy- and time-resolved two-photon
photoemission spectroscopy (2PPE) \cite{Hoefer98, Hoefer02,
Fauster99, Fauster95}. Theoretical understanding of the involved
electron-electron scattering processes has established that the
lifetime of image-potential-state electrons is determined by
interband scattering with bulk electrons and by intraband
contributions \cite{Hoefer02, Chulkov98}.

The high LDOS of the image-potential states near the surface makes
them accessible to STM. They appear at rather high sample bias
voltages of $4-8 \mathrm V$. The electric field between tip and
sample induces a Stark shift of the eigenenergies to higher
values. Interest in image-potential states modified by the
presence of an STM tip has focused so far on the fact that the
energetic positions of the states are sensitive to the electronic
structure of the surface \cite{Binnig85,Himpsel95}. Spectroscopy
of image-potential states by STM has been used to achieve chemical
contrast on the nanometer scale for metals on metal surfaces, e.g.
Cu on Mo(110) \cite{Himpsel95} or Fe/Cr surface alloys
\cite{Kuk99}.

In this letter, we present for the first time STM measurements of
the dynamical properties of the image-potential states in front of
the surface. Due to elastic scattering at point defects and step
edges of the electrons injected into these states, modulations of
the LDOS emerge. Using STS, we determine the energies for the
first four image-potential states as well as the dispersion
relation and the phase coherence length of the $n=1$ Stark-shifted
image-potential state on Cu(100). This opens up the possibility to
study these quantities locally in nanostructures. In the case of
surface state electrons at metal surfaces the local dynamics has
been studied by STS in great detail \cite{Li98a, Buergi99,
Kliewer00, Vitali03}. The results show a good agreement between
photoemission and STS experiments on the one side and theory on
the other side \cite{Kliewer00, Vitali03}. A similar comparison is
missing in the case of image-potential states.

We used a Cu(100) single crystal sample, carefully prepared by
sputtering and annealing cycles in UHV (base pressure $1\cdot
10^{-10}\mathrm{mbar}$). After cleaning the sample was transferred
{\it in situ} into a home-built STM operating at $6\mathrm K$.
Spectroscopic measurements were performed using a lock-in
technique with a modulation of the sample voltage of
$10-30\mathrm{mV_{RMS}}$ at a frequency of $4.5 \mathrm{kHz}$. All
bias voltages $V$ are sample potentials measured with respect to
the tip.

\begin{figure}[btp]
\includegraphics{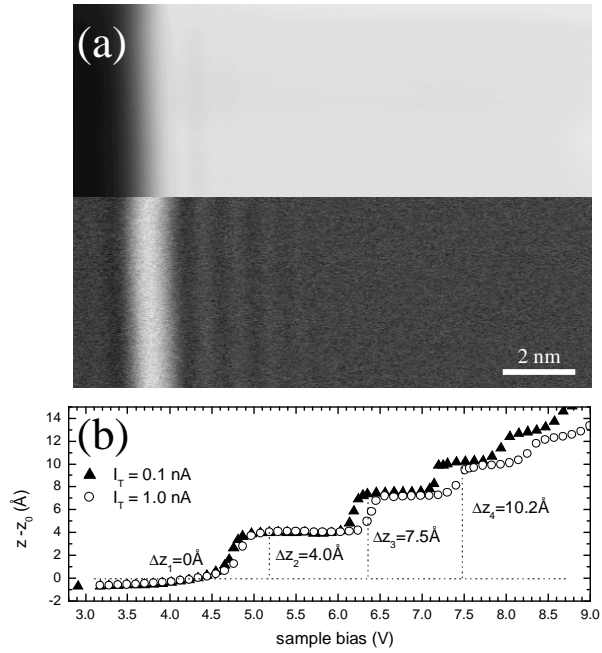}
 \caption[]{(a) Upper half: STM topography of an artificially
 created step on Cu(100) taken at $5.2\mathrm V$ bias voltage,
 lower half: $\mathrm dI/\mathrm dV$-map of the same place at
 the same bias voltage. The image-potential-state electrons reflected at
 the step edge create a density modulation which appears as a
 standing wave pattern in the $\mathrm dI/\mathrm dV$-map.
 (b) Measured $z(V)$ curve on a terrace of Cu(100) using two different
 tunneling currents of $0.1$ and $1.0\mathrm{nA}$. We
 observe up to four image-potential states in the electric field of the STM tip.
 The measured increments in the tip-sample distance
 are labelled $\Delta z_n$.}
 \label{fig0}
\end{figure}

Fig.~\ref{fig0}(a) shows a typical STM image of an artificially
created step edge. The upper half is a topographic image and the
lower half shows the simultaneously acquired $\mathrm dI/\mathrm
dV$-map taken at the same bias voltage. It reveals the quantum
interference pattern of the $n=1$ image-potential state created by
elastic scattering of the electrons injected by the tip into the
image-potential state. Circular standing waves were also observed
at point defects. (Not shown here.) We will later present a
detailed analysis of these patterns leading to the dispersion
relation and the momentum resolved linewidth of the $n=1$
Stark-shifted image-potential state.

The energies of the Stark-shifted image-potential states can be
measured using $z(V)$ spectroscopy as shown in Fig.~\ref{fig0}(b).
For this experiment the feedback loop is kept active while
sweeping the bias voltage. To maintain a constant current the tip
is retracted with increasing bias voltage as more and more states
become available to the tunneling electrons. We have taken
$z(V)$-spectra at currents of $0.1$ to $1\mathrm{nA}$. The spectra
show a series of steps, where each step is due to the contribution
of a new image-potential state to the tunneling current allowing
us to identify the first four image-potential states. Their
energies $E_n$ relative to the vacuum level $E_\mathrm{vac}$ of
the sample are obtained from the bias voltages $V_n$, at which the
steps occur by
\def\mustequal{\mathop{\displaystyle{\raise6pt\hbox{!}{\mkern-15mu}=}}}
\begin{equation}
E_n=eV_n-\phi \label{eq1}
\end{equation}
where $\phi=4.6\mathrm{eV}$ \cite{Fauster95} is the work function
of Cu(100) and $e$ the elementary charge. Note that the states
appear above the vacuum level of the substrate but nevertheless
they are bound in $z$-direction by the tip and crystal potential
respectively. From the absence of any features in $z(V)$ below
$3\mathrm{V}$ we identify the step at $V=4.7\mathrm{V}$ as the
state $n=1$. The energies are considerably larger compared to the
unperturbed states which form a Rydberg series below
$E_\mathrm{vac}$ starting at $-0.6\mathrm{eV}$\cite{Chulkov99,
Hoefer02}.

As can be seen from a comparison of the two spectra shown in
Fig.~\ref{fig0}(b), the states shift to higher energies for higher
tunneling currents due to the decreased tip-sample distance.
Measurements with different tips, i.e. tips that have been
modified by field emission and gentle dipping into the surface,
reveal a dependence of the energy levels on the tip properties
which is stronger for the higher states. While the $n=1$ state
remains at a bias voltage of $4.7\pm 0.1\mathrm V$, we observe the
higher states to shift by as much as $0.5\mathrm V$ ($n=2$) and
$0.9\mathrm V$ ($n=3$). This yields a much stronger dependence of
the energy levels on tip properties than on tunneling conditions,
i.e the current at which the $z(V)$ spectroscopy is performed.

\begin{figure}[btp]
 \includegraphics{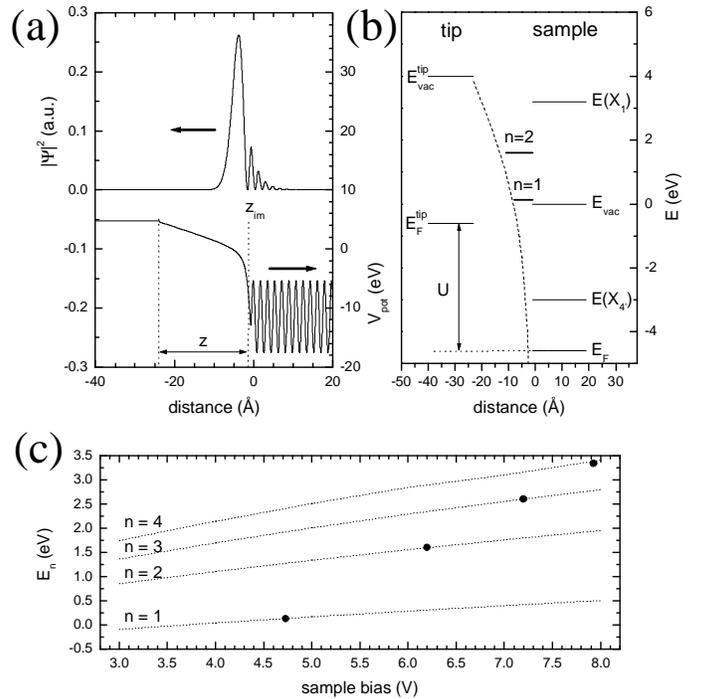}
 \caption[]{ (a) one-dimensional model potential used and probability density
 of the wave function of the first image-potential state obtained.
 (b) schematic drawing of the energy levels of the
 resonances, the Fermi levels of tip and sample, the bulk band
 edges of the Cu(100) crystal ($E(X_{4^\prime})$ and $E(X_1)$) and
 the potential in the tunneling gap (dashed line).
 (c) calculation (dotted lines) of the energies
 of the image-potential states in the electric field of the STM tip as function
 of the applied voltage. Each of the dotted lines corresponds to
 one particular state ($n=1$ to $4$) calculated for a tip-sample distance
 $z=\Delta z_n+z_0$, where $\Delta z_n$ is taken from the
 measurement shown in Fig.~\ref{fig0}(b) and $z_0=22.5\mathrm\AA$. Full
 circles are plotted at the voltages where the steps in Fig.~\ref{fig0}(b) occur
 and at the corresponding energies according to Eq.~\ref{eq1}.}
 \label{fig1}
\end{figure}

To understand these findings we performed model calculations using
a one-dimensional potential as introduced by Chulkov et al.
\cite{Chulkov99}. This potential reproduces the Rydberg series of
the image-potential states and the positions of the projected band
edges at the $\overline{\Gamma}$ point ($k_{||}=0$) in the Cu(100)
surface Brillouin zone. We integrated the Schr\"{o}dinger equation
in real space employing the model potential for a 25 layer
crystal. The influence of the tip is modelled by adding as
function of the bias voltage a linearly increasing potential to
the image potential of the crystal reaching from the point
$z_\mathrm{im}$ (see Ref.~\onlinecite{Chulkov99}) to the point
$z_\mathrm{im}+z$ where $z$ is the tip-sample distance. A similar
ansatz has been used in Ref.~\onlinecite{Limot03}. Since the
change in tip-sample distance is given by the plateaus $\Delta
z_n$ (Fig. ~\ref{fig0}(b)) we use $z=\Delta z_n+z_0$ and treat
$z_0$ as the only adjustable parameter in the model assuming equal
work functions of tip and sample. This choice was made in favor of
discussing the average electric field in the junction
\cite{Binnig85} since it allows to separate $V$ as externally
controllable parameter from details of the potential. The model
potential and the probability distribution of the resulting wave
function are shown in Fig.~\ref{fig1}(a). Fig~\ref{fig1}(b) shows
schematically the resulting energy level diagram.  In such a
simple model, the energies $E_n$ of the image-potential states in
the electric field of the STM tip are reproduced for all $n$
observed. This is shown in Fig.~\ref{fig1}(c), where the
calculated energies $E_n$ are plotted as a function of the applied
bias voltage for $z_0=22.5\mathrm\AA$ corresponding to the
measurement with a tunneling current of $0.1\mathrm{nA}$. The
agreement is excellent. To reproduce the energy levels for the
measurement at a current of $1\mathrm{nA}$ a $z_0=19.5\mathrm\AA$
is found. We emphasize that there is no need for an n-dependent
"surface-corrugation parameter" as was employed earlier
\cite{Binnig85}. To arrive at the expected smaller $z_0$ values of
$8$ to $10\mathrm\AA$ \cite{Gim87} one has to improve the
treatment of the tip electrode. The detailed inclusion of the
image potential at the surface of the tip in a calculation using
two Cu(100) model potentials facing each other yielded $z_0$
values which were $3-5\mathrm\AA$ lower than the ones found above.
On the other hand, the radius of curvature $R$ of the tip can be
neglected. Only for unrealistically sharp tips with $R\approx z$
the potential near the tip will fall off appreciably more quickly
than the linear potential. We conclude that to explain the
variation of the $E_n$ with different tips the contact potential
and not the tip radius is the decisive quantity. From the
experimental point of view it is quite likely that the tungsten
tip is coated with copper, since it is frequently prepared by
slightly dipping it into the surface. Both, the composition and
the morphology of the very end of the tip can lead to a lower work
function compared to that of the Cu(100) surface, which can be
compensated by a reduced $z_0$.

\begin{figure}[btp]
\includegraphics{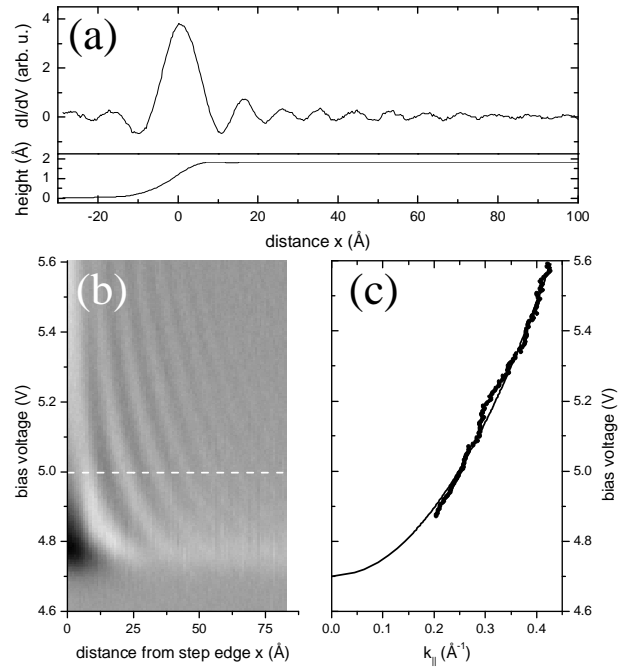}
 \caption[]{
 (a) $\mathrm dI/\mathrm dV$ signal at 5.0V sample bias (upper graph) and
 topographic signal at a step edge. Electron interference produces
 the oscillations in $\mathrm dI/\mathrm dV$.
(b) $\mathrm dI/\mathrm dV(V,x)$-map. $\mathrm dI/\mathrm dV$ is
plotted as a function of the lateral distance $x$ from a step edge
and of the bias voltage $V$ \cite{destripe}. Horizontal line cuts
reveal the standing waves due to the interference of electrons in
the image-potential state scattered at the step edge. The dashed
line indicates the energetic position of the data shown in (a).
(c) points represent the $k$-values obtained from a fit to each
line in (b), the solid line shows the parabolic dispersion fitted
to these points.}
 \label{fig2}
\end{figure}

In the following, the dynamics of the the image-potential-state
electrons in front of the surface will be discussed in detail. Due
to elastic scattering of these electrons at point defects and step
edges, modulations of the LDOS are created through quantum
interference (Fig.~\ref{fig0}(a) and Fig.~\ref{fig2}(a)). This
allows to study the dynamics of the states with non-vanishing
momentum parallel to the surface locally. The analysis of the
interference pattern of electrons scattered at a step edge enables
the determination of their wave vector and phase coherence length
as a function of energy. The interference pattern is measured
through the $\mathrm dI/\mathrm dV$ signal which is proportional
to the LDOS at the given energy. In Fig.~\ref{fig2}(b) $\mathrm
dI/\mathrm dV (V,x)$ is measured for bias voltages ranging from
$4.6\mathrm V$ to $5.6\mathrm V$ at increasing distances $x$ from
the step edge. The resulting curves are represented as a grey
scale map, where horizontal line sections are the energy resolved
electron density oscillations as shown in Fig.~\ref{fig2}(a). The
density oscillations reveal the parabolic dispersion relation of
the state with $E_{\overline{\Gamma}}^\prime=4.7\pm0.1\mathrm{eV}$
and $m^\prime=(0.8\pm0.1)m_0$ (Fig~\ref{fig2}(c)). However, with
the help of the calculations presented above, the influence of the
tip on the dispersion of the image-potential state can be
corrected for. Since the data are collected in open feedback mode,
i.e. the distance between tip and sample is kept constant, one
needs only to compensate for the shift of the energy at
$\overline{\Gamma}$ with changing electric field during the bias
voltage sweep. The dependence of the state's energy on the applied
bias voltage close to $E_{\overline{\Gamma}}^\prime$ is
approximately linear. From the calculations shown in
Fig.~\ref{fig1}(c) we get $\mathrm dE_1/\mathrm dV = 0.12$. Using
this correction, we obtain an effective mass of
$m^*=m^\prime/(1-0.12)=(0.9\pm0.1)m_0$. This agrees perfectly with
the effective mass $m^*=0.9m_0$ of the $n=1$ image-potential state
as determined by 2PPE \cite{Fauster95}. Similarly we find an
effective mass of the $n=2$ state of $m^*=(1.0\pm0.1)m_0$. An
increase in the effective mass for $n=2$ was also observed by 2PPE
\cite{Weinelt02}.

\begin{figure}
\includegraphics{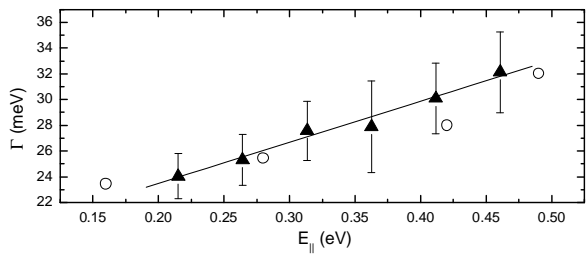}
 \caption[]{Phase coherence lengths of the interference pattern at a step
 edge converted into linewidths $\Gamma(E)$. Solid symbols are the data taken by STM, the solid line is a linear fit to
 them; open symbols represent linewidths determined by 2PPE measurements from Ref. \onlinecite{Hoefer02}.}
 \label{fig3}
\end{figure}

The standing wave pattern decays with increasing distance from the
step edge due to geometric factors and to a loss of coherence
\cite{Buergi99}. The decaying LDOS pattern formed by the
image-potential states near a step edge can be described by
\begin{equation}
\rho\approx L_0\left(1-re^{-2x/L_\Phi}{\mathrm
J}_0(2k_{||}x)\right) \label{decay}
\end{equation}
where $L_0$ is an overall proportionality constant, $r$ the
reflectivity of the step edge, $L_\Phi$ the phase coherence
length, $k_{||}$ the wave vector of the electron parallel to the
surface plane, and ${\mathrm J}_0$ the Bessel function of zeroth
order.

To determine the lifetime of an electron in the $n=1$
image-potential state we have analyzed quantitatively the quantum
interference pattern and measured the phase coherence length
$L_\phi$ as a function of energy on large defect free terraces.
The scattering processes thus studied are those experienced in the
absence of any defects \cite{Buergi99, Vitali03}. Inelastic
scattering at the step edge leads to a reduced overall amplitude
of the standing waves described by $r$ in Eq.~\ref{decay}. Care
was taken to account for instrumental broadening through the
applied bias modulation when measuring the $\mathrm dI/\mathrm dV$
signal which also induces a decay of the wave pattern
\cite{Buergi99}. The obtained phase coherence lengths of
$75-85\mathrm \AA$ are converted into linewidths through
$\Gamma(E) = \hbar^2 k/(m^* \cdot L_\phi(E))$ by using the
measured $k$ and $m^*$ according to the dispersion relation
(Fig.~\ref{fig2}). The results shown in Fig.~\ref{fig3} agree
excellently with the $k$-resolved linewidths found by 2PPE
measurements \cite{Hoefer02}. In agreement with theory, we find
$\Gamma(E)$ to increase linearly with energy, although the rate of
$\mathrm d\Gamma/\mathrm dE = 32 \mathrm{meV}/\mathrm{eV}$
obtained from the fit in Fig.~\ref{fig3} is lower than the
theoretical prediction \cite{Hoefer02}. The comparison of our
results with the 2PPE measurements demonstrates that the presence
of the tip does not alter substantially the dynamical properties
of electrons in the image-potential states. Although the $n=1$
state shifts by as much as $0.7\mathrm{eV}$ due to the presence of
the electric field, it is still located near the center of the
$\sim6\mathrm{eV}$ wide directional band gap of Cu(100). There is
thus no significant change in the coupling to the bulk electrons,
which is the main contribution to the linewidth.

In conclusion, we observed for the first time density modulations
of image-potential-state electrons in the vicinity of steps and
point defects on Cu(100). The quantum interference patterns allow
us to determine the dispersion relation of the states which
experience a Stark shift due to the electric field in the
tunneling junction. Furthermore we measured the phase coherence
length of the electrons on clean terraces. Both, the dispersion
relation and the phase coherence length agree well with the
results of non-local photoemission experiments. This shows that
the motion of the electrons parallel to the surface is not
noticeably affected by the field of the STM tip. Due to the local
character of the measurement the STM can therefore be used to
study the dynamical behavior of image-potential-state electrons
confined laterally to nanostructures and to characterize the
scattering properties of surface defects and adsorbates.


\begin{references}
\bibitem{destripe} To enhance the contrast, the average value of each horizontal line has been subtracted.
\end{references}
\end{document}